# Pneumatically-Controlled Tactile Actuating Modules for Enhanced VR Safety Training

Ahsan Raza*, Mohammad Shadman Hashem*, and Seokhee Jeon*

*Department of Computer Engineering, Kyung Hee University, Yongin-si, South Korea*

*(Email: jeon@khu.ac.kr)*

**Abstract ---** Our system introduces a modularized pneumatic actuating unit capable of delivering vibration, pressure, and impact feedback. Designed for adaptability, these modular tactile actuating units can be rapidly customized and reconfigured to suit a wide range of virtual reality (VR) scenarios, with a particular emphasis on safety training applications. This flexibility is demonstrated through scenarios such as using construction tools in a virtual environment and simulating safety protocols against falling objects. Innovative mounting solutions securely attach the actuators to various body sites, ensuring both comfort and stability during use. Our approach enables seamless integration into diverse VR safety training programs, enhancing the realism and effectiveness of simulations with precise and reliable haptic feedback.

## 1 INTRODUCTION

Haptic feedback is crucial for enhancing interactive experiences in Virtual Reality (VR), significantly deepening immersion by realistically simulating tactile interactions [1]. As technology rapidly advances, our communication and interaction modes evolve, moving from simple telephonic communications to complex, immersive VR scenarios. This shift makes experiences more engaging and lifelike, expanding the horizons of digital interaction. Augmenting VR's visual capabilities with haptic feedback enhances realism, making virtual environments more believable and physically palpable [2]. In haptic technology, interactions are classified into two main types: tactile feedback, which utilizes the skin's sensitivity to vibrations, textures, and pressure, and kinesthetic feedback, which targets muscles, joints, and tendons with force or torque feedback to simulate handling real objects. This study focuses primarily on modular tactile feedback interface, exploring its applications across various modalities to enhance the immersion of virtual reality (VR) experiences in training, educational tools, and entertainment. The goal is to increase user engagement and improve learning outcomes by providing a sensory experience that closely mimics real-world interactions

The concept of modularity in haptic interfaces entails the use of customizable haptic modules designed to deliver specific types of tactile feedback tailored for diverse applications. Recent studies have developed modular actuating solutions capable of generating multimodal tactile feedback using either single or multiple actuators for various uses [3], [4], [5]. Although the modular nature of these solutions broadens their applicability, a frequent issue arises with their ability to provide a variety of feedback within a single module [6], often leading to bulkier systems [7].

To address these challenges, we introduce an advanced tactile actuation module capable of delivering multimodal feedback, including vibration, pressure, and impact, in a compact form factor. This actuator is designed to be adaptable to multiple body sites thanks to its innovative design and mounting mechanism. An illustration of the proposed system is shown in Figure 1 and 2. The actuation mechanism is pneumatically powered and controlled through a hardware setup that includes solenoid valves. These valves precisely regulate the air pressure within the actuator's chamber, allowing for the modulation of tactile feedback to replicate diverse environmental interactions. This configuration not only facilitates precise sensation tuning but also ensures rapid response adjustments to meet the dynamic simulation demands faced by users.

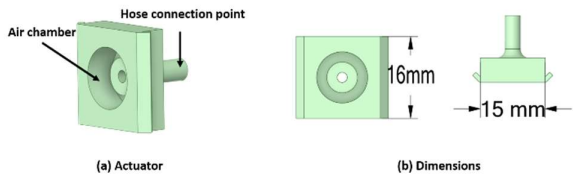

Figure 1: Actuator's 3D model along with the dimensions

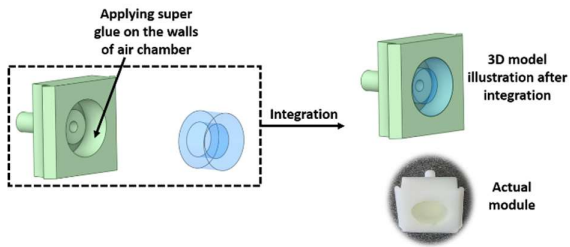

Figure 2: Illustration of the integration of the actual actuator.

## 2 Demonstration

In the first demonstration scenario, our system is deployed in a VR setup designed to simulate the environment of an earthquake, where participants experience sensations of falling debris. This setup is critical for training individuals on how to react during such emergencies, with actuating units strategically placed on the head to simulate the impact of falling objects. This not only enhances realism but also reinforces safety protocols by providing physical feedback alongside visual cues, creating a cohesive and immersive experience. The second scenario focuses on operating machinery and tools generally utilizes for construction work within a virtual setting, essential for industrial safety training. Here, tactile units replicate the vibrations and pressures of operating heavy machinery or power tools, like drills or chainsaws, through the pneumatic system, providing real-time feedback that enhances the operator's understanding and handling of the tool. This allows users to experience the physical feedback of operating various tools without real-world risks, with feedback intensity adjusting in real-time to mirror different operational states of the tools thus enhancing training effectiveness. Furthermore, the modular nature of the tactile units facilitates easy integration into existing VR systems and quick reconfiguration for different scenarios, making them ideal for varied training modules in industrial and emergency response applications. The system's adaptability also extends to its mounting solutions, which are innovatively designed to ensure the actuators remain securely attached to the user, providing consistent feedback without sacrificing comfort.

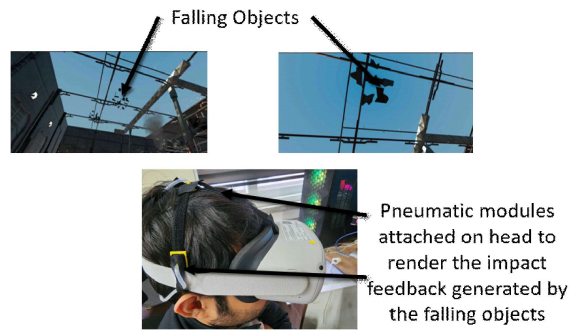

Figure 3: An illustration of falling objects in the VR scene.

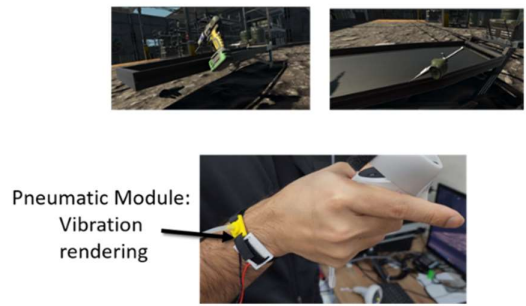

Figure 4: An illustration of operating tool in the VR environment.

## 3 Conclusion

This work presents the transformative potential of modular haptic feedback systems in Virtual Reality (VR), enhancing immersive experiences across diverse applications. Advanced tactile actuation modules deliver varied sensory feedback, effectively bridging the gap between digital and real-world interactions. Demonstration scenarios, from emergency responses to industrial settings, underscore the systems' adaptability and realism. The modularity ensures seamless integration into various VR platforms, optimizing user engagement and training effectiveness. Innovative mounting solutions maintain user comfort and feedback consistency, promising further advancements in interactive technology applications.

ACKNOWLEDGEMENT

This research was supported by the IITP under the Ministry of Science and ICT Korea through the IITP program No. 2022-0-01005 and under the metaverse support program to nurture the best talents (IITP-2024-RS-2024-00425383).

REFERENCES

[1] GN. Magnenat-Thalmann and U. Bonanni, "Haptics in virtual reality and multimedia," IEEE MultiMedia, vol. 13, no. 3, pp. 6–11, 2006.

[2] F. Danieau, A. L´ecuyer, P. Guillotel, J. Fleureau, N. Mollet, and M. Christie, "Enhancing audiovisual experience with haptic feedback: a survey on hav," IEEE transactions on aptics, vol. 6, no. 2, pp. 193–205, 2012.

[3] G. Park, H. Cha, and S. Choi, "Haptic enchanters: Attachable and detachable vibrotactile modules and their advantages," IEEE transactions on haptics, vol. 12, no. 1, pp. 43–55,

[4] B. Zhang and M. Sra, "Pneumod: A modular haptic device with localized pressure and thermal feedback," in Proceedings of the 27th ACM Symposium on Virtual Reality Software and Technology, 2021, pp. 1–7.

[5] Raza, A., Hassan, W. and Jeon, S.: "Pneumatically Controlled Wearable Tactile Actuator for Multi-Modal Haptic Feedback," IEEE Access (2024).

[6] R. E. S. Cruz, M. C. Coffey, A. Y. Sawaya, and R. P. Khurshid, "Modular haptic feedback for rapid prototyping of tactile displays," in 2021 IEEE World Haptics Conference (WHC).IEEE, 2021, pp. 703–708.

[7] R. Zhou, Z. Schwemler, A. Baweja, H. Sareen, C. L. Hunt, and D. Leithinger, "Tactorbots: a haptic design toolkit for out-of-lab exploration of emotional robotic touch," in Proceedings of the 2023 CHI Conference on Human Factors in Computing Systems, 2023, pp. 1–19.